\documentclass[fleqn,usenatbib]{mnras}

\usepackage[T1]{fontenc}
\usepackage{physics}
\usepackage{tabularx}
\DeclareRobustCommand{\VAN}[3]{#2}
\let\VANthebibliography\thebibliography
\def\thebibliography{\DeclareRobustCommand{\VAN}[3]{##3}\VANthebibliography}

\usepackage{graphicx}	
\usepackage{amsmath}	
\usepackage{amssymb}	
\usepackage{subcaption}
\usepackage{newtxtext,newtxmath}

\title[Non-thermal emission from tidal ejecta]{Late-time non-thermal emission from mildly relativistic tidal ejecta of compact objects merger}

\author[Sadeh]{
Gilad Sadeh\thanks{E-mail: gilad.sade@weizmann.ac.il}
\\
\it{Dept. of Particle Phys. \& Astrophys., Weizmann Institute of Science, Rehovot 76100, Israel}
}

\date{Accepted XXX. Received YYY; in original form ZZZ}

\pubyear{2024}

\begin{document}
\label{firstpage}
\pagerange{\pageref{firstpage}--\pageref{lastpage}}
\maketitle

\begin{abstract}
Mergers of compact objects (binary neutron stars, BNS, or neutron star-black hole, NSBH) with a substantial mass ratio ($q>1.5$) are expected to produce a mildly relativistic ejecta within $\sim20^\circ$ from the equatorial plane. We present a semi-analytic approach to calculate the expected synchrotron emission observed from various viewing angles, along with the corresponding radio maps, that are produced by a collisionless shock driven by such ejecta into the interstellar medium. 
This method reproduces well (up to $\sim30\%$ deviations) the observed emission produced by 2D numerical calculations of the full relativistic hydrodynamics. 
We consider a toroidal ejecta with an opening angle of $15^\circ\leq\theta_
\text{open}\leq30^\circ$ and broken power-law mass distribution, $M(>\gamma\beta)\propto(\gamma\beta)^{-s}$ with $s=s_{\rm KN}$ at $\gamma\beta<\gamma_0\beta_0$ and $s=s_{\rm ft}$ at $\gamma\beta>\gamma_0\beta_0$ (where $\gamma$ is the Lorentz factor). The parameter values are chosen to characterize merger calculation results- a "shallow" mass distribution, $1<s_{\rm KN}<3$, for the bulk of the ejecta (at $\gamma\beta\approx 0.2$), and a steep, $s_{\rm ft}>5$, "fast tail" mass distribution.
While the peak flux is dimmer by a factor of $\sim$2-3, and the peak time remains roughly the same (within $20\%$), for various viewing angles compared to isotropic equivalent ejecta ($\theta_\text{open}=90^\circ$) considered in preceding papers, the radio maps are significantly different from the spherical case. 
The semi-analytic method can provide information on the ejecta geometry and viewing angle from future radio map observations and, consequently, constrain the ejection mechanism. For NSBH mergers with a significant mass ejection ($\sim0.1M_\odot$), this late non-thermal signal can be observed to distances of $\lesssim 200$Mpc for typical parameter values.
\end{abstract}




\section{Introduction}
Binary mergers involving compact objects with a substantial mass ratio ($q>1.5$) or ones in which a prompt black hole (BH) is produced, such as NSBH \citep{rosswog_mergers_2005,just_comprehensive_2015,kawaguchi_black_2015,kyutoku_neutrino_2018,foucart_dynamical_2017,foucart_brief_2020} or BNS \citep{radice_binary_2018,bernuzzi_accretion-induced_2020,nedora_numerical_2021} systems, yield expanding ejecta that attains mildly relativistic velocities, $\gamma\beta>0.1$, through tidal interaction between the BH/NS and the NS \citep[see][for review]{bernuzzi_neutron_2020,kyutoku_coalescence_2021,chen_black_2024}. Such ejecta is expected to produce non-thermal electromagnetic (EM) emission on a time scale of $\sim$10 years \citep{nakar_radio_2011,sadeh_non-thermal_2023}. 
BNS and NSBH mergers are likely the sources of (short) gamma-ray bursts \citep[see][for reviews]{meszaros_theories_2002,piran_physics_2004,nakar_short-hard_2007}. Consequently, these events are expected to produce highly relativistic jets that may dominate the non-thermal emission from the tidally driven ejecta, depending on the jet's energy, opening angle, and the angle at which it is observed. Typically, because of the jet's lower mass and higher Lorentz factor, the late-time emission is expected to be dominated by the tidal ejecta. To interpret future observations accurately, it will be necessary to disentangle these two components.

NSBH mergers can exhibit disruptive or non-disruptive behaviors. In the former, the NS is disrupted by the tidal field of the BH, triggering mass ejection, formation of a disk around the BH, and the emission of EM radiation. 
A disruptive NSBH merger occurs when the disruption radius, $R_\text{dis}$, at which the tidal forces of the BH are larger than the gravitational forces that hold the NS as a compact object, is larger than the radius of the innermost stable circular orbit, $R_\text{ISCO}$.
Conversely, in non-disruptive mergers, the NS plunges as a whole into the BH, and the phenomenon is detectable solely through gravitational wave (GW) measurements.
Numerical relativity (NR) simulations of BNS mergers suggest that the mass ratio between the BNS significantly affects the ejecta geometry \citep{bernuzzi_accretion-induced_2020,bernuzzi_long-lived_2024}. BNS mergers of comparable mass ($q\sim 1$) eject mass dynamically due to shocks driven by the collision and also due to tidal forces. The ejecta geometry for such mergers is quasi-spherical \citep{radice_binary_2018,nedora_numerical_2021}.
Additional spiral-wave wind component \citep{nedora_numerical_2021,radice_secular_2023} should be considered in case of long-lived neutron star remnant (absent for NSBH systems). 
In cases where the mass ratio is considerable, $q>1.5$, the ejected mass is concentrated within $20^\circ$ from the equatorial plane since most of the mass is ejected by the tidal forces.
In BNS mergers in which a prompt BH is produced, the dynamical ejecta is also dominated by the tidal component. Correspondingly, the angular spread of the ejecta beyond the equatorial plane is within $\sim20^\circ$ \citep{radice_binary_2018,nedora_numerical_2021}.
In all cases the NR simulations suggest dynamical ejecta of $\sim$ $10^{-3}-10^{-2}M_\odot$, for various equation of states (EoSs) and mass ratios \citep[see][]{radice_dynamics_2020}, with velocity profile that is well approximated by the one suggested in \citet{sadeh_non-thermal_2023} and confirmed in \citet{zappa_binary_2023}.
NSBH mergers with comparable mass have not been observed so far and, thus, have not been considered in NR simulations.
NSBH merger simulations of large mass ratio, $q>3$, provide some constraints over the dynamical ejecta of such events \citep{rosswog_mergers_2005,just_comprehensive_2015,kyutoku_dynamical_2015,kyutoku_neutrino_2018,foucart_black-hole--neutron-star_2013,foucart_dynamical_2017,chen_black_2024}. In cases of tidal disruption, they imply the dynamical ejecta has only a tidal component of $\sim$ $10^{-2}-10^{-1}M_\odot$, for various initial parameters and EoSs. 
This ejecta shares a similar geometry to BNS dynamical ejecta with a high mass ratio according to numerical relativity simulations \citep{bernuzzi_accretion-induced_2020,bernuzzi_long-lived_2024}; furthermore, it is consistent with the velocity profile suggested in \cite{sadeh_non-thermal_2023} for BNS mergers \citep{kyutoku_dynamical_2015,chen_black_2024}.
The amount of ejected mass, its geometry, and its velocity distribution depend on the parameters of the merging system (whether it is NSBH or BNS), such as the mass and the spin of the BH (in the case of NSBH), the mass of the NS, the nuclear EoS and the mass ratio, $q$ \citep{rosswog_mergers_2005,shibata_merger_2008,lovelace_massive_2013,kyutoku_dynamical_2015,foucart_dynamical_2017,hayashi_general-relativistic_2022,chen_black_2024}.
For example, for a mass ratio of $q>8$, no considerable mass ejection is expected since the NS is not tidally disrupted \citep{foucart_dynamical_2017}. 
Measurements of the EM emission generated by the expanding ejecta will enable us to constrain its properties and, hence, the abovementioned parameters.
The opening angle of the ejecta is approximately conserved \citep{kyutoku_dynamical_2015} because the velocity direction does not change appreciably once the hydrodynamic merger interaction becomes negligible.
The azimuthal opening angle of the ejecta varies from $180^\circ$ to $360^\circ$ for different initial conditions \citep{kyutoku_dynamical_2015,kawaguchi_models_2016,bernuzzi_accretion-induced_2020}.
\cite{margalit_radio_2015} consider the effect of a non-spherical (and non-relativistic) ejecta structure on the synchrotron light curve. They provide a rough approximation of the peak time scale that is delayed with respect to the isotropic equivalent case and reach the conclusion that the peak flux scale remains roughly the same as in the isotropic equivalent case. However, they do not provide full numerical calculations and base their conclusions on an analytic approximation from \cite{nakar_radio_2011}.
It should be noted here that more material is expected to be ejected from the remnant disk \citep{fernandez_dynamics_2017}. Due to the lower velocities and time delay compared to the dynamical ejecta, no mixing between the two components is expected \citep{fernandez_interplay_2015}. Furthermore, since the non-thermal signal is highly sensitive to the shocked plasma Lorentz factor \citep{sadeh_non-thermal_2023}, the lower velocity of the disk outflow \citep{fernandez_dynamics_2017,fernandez_long-term_2019} would make its potential non-thermal signal much less bright.

The rate of BNS mergers was estimated by the Galactic BNS systems to be $\sim 10\left(\frac{D}{200\text{Mpc}}\right)^3\text{yr}^{-1}$ \citep{phinney_rate_1991,pol_future_2019,pol_updated_2020}. Such estimation is not possible for the rate of NSBH mergers since our Galaxy has no known NSBH binary systems. Due to the low number of detected BNS and NSBH mergers via GW, the merger rate obtained from these observations is extremely uncertain. For BNS mergers it is between $0.1$ to $14\left(\frac{D}{200\text{Mpc}}\right)^3\text{yr}^{-1}$ and for NSBH mergers it is between $0.1$ to $1.5\left(\frac{D}{200\text{Mpc}}\right)^3\text{yr}^{-1}$ \citep{ligo_scientific_collaboration_population_2023,the_ligo_scientific_collaboration_observation_2024}.
The fraction of EM bright events out of the GW detectable NSBH mergers events is currently unknown due to the low statistics- only three confirmed events so far in the third and fourth LIGO-Virgo-KAGRA run \citep{abbott_observation_2021,the_ligo_scientific_collaboration_observation_2024}. However, first attempts have been made \citep{fragione_black-hole-neutron-star_2021,biscoveanu_population_2023,the_ligo_scientific_collaboration_observation_2024}, suggesting a fraction of $\sim10-30\%$. \cite{martineau_black_2024} estimated the ejected mass in the observed event, GW230529, to be negligible, $\sim10^{-5}M_\odot-10^{-3}M_\odot$, due to the low BH spin $\chi_\text{BH}\lesssim 0.1$.
While the mass ratio in NSBH mergers is likely to be above $q=1.5$ considered the observed mergers so far, for the observed BNS mergers GW170817 and GW190425, the mass ratio estimate varies between $q\approx1$ to $q\approx2.5$ for different spin priors. 

This paper explores synchrotron emission spanning the radio-to-X-ray range, originating from a collisionless shock driven by mildly relativistic toroidal ejecta with an opening angle of $15^\circ$ to $30^\circ$ and a broken power-law dependence of mass on momentum, with parameter values characteristic of the results of numerical calculations of the BNS ejecta (see \S \ref{sec:geometry}). The ejecta expands into a uniform interstellar medium (ISM) characterized by a number density $n$. We derive a semi-analytic calculation method for the case of axisymmetric distribution of mass \citep[as in][]{gompertz_multi-messenger_2023} exhibiting a power-law momentum-mass dependence \citep{sadeh_non-thermal_2023}, where the parameter values are consistent with the results of numerical calculations of the tidal ejecta 
\citep{kyutoku_dynamical_2015,kyutoku_neutrino_2018,foucart_dynamical_2017,bernuzzi_accretion-induced_2020,nedora_numerical_2021,chen_black_2024,bernuzzi_long-lived_2024}. This semi-analytic method is valid for as long as the reverse shock did not cross the fast tail part of the ejecta, during this phase the shocked plasma is not expected to go through a significant lateral expansion, so we assume angular independent distribution of shocked plasma parameters.
We derive the non-thermal emission spectra under the assumption of a power-law energy distribution for electrons, given by $dn_e/d\gamma_e\propto \gamma_e^{-p}$. Here, $\gamma_e$ represents the electron Lorentz factor within the plasma's rest frame, where $p$ is within the range of $2\leq p\leq2.5$. Concurrently, the fractions $\varepsilon_e$ and $\varepsilon_B$ correspond to the post-shock internal energy density proportions attributed to non-thermal electrons and magnetic fields, respectively. This phenomenological description, capturing the post-shock plasma conditions, finds support across a diverse spectrum of observations and plasma calculations encompassing both relativistic \citep[see][]{keshet_energy_2005,waxman_gamma-ray_2006,keshet_analytical_2006,bykov_fundamentals_2011,sironi_maximum_2013,kobzar_electron_2021} and non-relativistic shocks \citep[see][]{blandford_particle_1987, pohl_pic_2020,ligorini_mildly_2021}.
We verified the accuracy of the semi-analytic results for the non-thermal emission presented in this paper through a comprehensive comparison with full 2D numerical computations across a wide range of parameter configurations. 

The paper is organized as follows. In \S~\ref{sec:geometry}, we describe the framework of the calculations presented later on.
We derive the semi-analytic calculation for the synchrotron light curve and corresponding radio map in \S~\ref{sec:semi}.
Then, we describe the 2D relativistic-hydrodynamics numerical calculation and the post-processing we use to produce light curves in \S~\ref{sec:numerical}.
The accuracy of the semi-analytic method is estimated by comparison with the detailed numeric calculations in \S~\ref{sec:comparison}, while in \S~\ref{sec:results}, we provide a few examples for light curves and radio maps produced by the semi-analytic approach. Finally, our conclusions are summarized in \S~\ref{sec:conclusions}.

\section{Calculation framework}
\label{sec:geometry}
\subsection{Ejecta geometry description}
\begin{figure}
    \includegraphics[width=9cm]{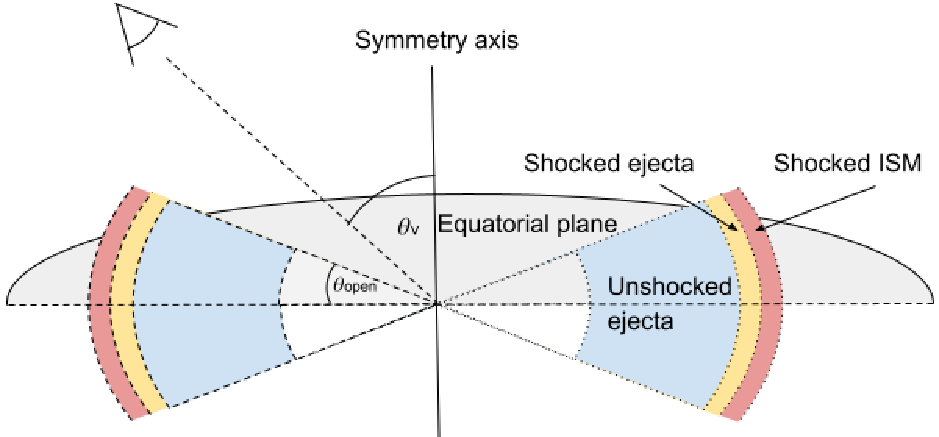}
    \caption{A schematic illustration showing a 2D slice of the 3D toroid ejecta (described in \S~\ref{sec:geometry}), and the forward-reverse shock structure that is formed and described in \S~\ref{sec:dynamics}.
    The symmetry axis corresponds to the azimuthal symmetry of toroids. 
    $\theta_\text{open}$ is the angle between the equatorial plane and the initial ejecta opening angle. $\theta_\text{v}$ is the angle between the symmetry axis and the line of sight, such that $\theta_\text{v}=\pi/2$ corresponds to the equatorial plane.
    }
    \label{fig:geometry}
\end{figure}
\label{sec:analytic}
In \cite{sadeh_non-thermal_2023,sadeh_non-thermal_2024-1}, we considered a spherical ejecta with the following mass profile 
\begin{equation}
\label{eq:profile}
    M(>\gamma\beta)= M_0
    \begin{cases}
    \left(\frac{\gamma\beta}{\gamma_0\beta_0}\right)^{-s_\text{ft}} & \text{for}\quad \gamma_0\beta_0<\gamma\beta,\\
    \left(\frac{\gamma\beta}{\gamma_0\beta_0}\right)^{-s_\text{KN}} & \text{for}\quad 0.1<\gamma\beta<\gamma_0\beta_0,
    \end{cases}    
\end{equation}
with parameter values characteristic of the results of numerical calculations of the BNS ejecta; $0.3<\beta_0<0.5$, $5<s_\text{ft}<12$, $0.5<s_\text{KN}<3$, and $10^{-6}M_\odot<M_0<10^{-4}M_\odot$. This analytic form provides a good approximation for the variety of ejecta profiles obtained in NR simulations of BNS mergers \citep{zappa_binary_2023}.
A merger of compact objects with a mass ratio of $q>1.5$, or one in which a prompt BH is produced, is expected to eject mass mainly within an angle of $\sim20^\circ$ from the equatorial plane and follow the same velocity profile as the dynamical ejecta of mergers with comparable masses \citep{kyutoku_dynamical_2015,foucart_dynamical_2017,radice_binary_2018,bernuzzi_accretion-induced_2020,nedora_numerical_2021,bernuzzi_long-lived_2024}. 
 The ejecta azimuthal structure varies from full azimuthal symmetry to a crescent shape with an azimuthal opening of $\phi\approx 180^\circ$ \citep{kyutoku_dynamical_2015}.
The geometric structure we consider to describe such ejecta is a toroid (sphere truncated by a double-sided cone), which limits us to the case of full azimuthal symmetry, with a mass profile that is given by Eq.~{\ref{eq:profile}}, and various opening angles, $15^\circ-30^\circ$, that are consistent with the results of NR simulations, at some initial lab time $t_0$ (rest frame of the ISM). The angle between the equatorial plane and the ejecta "edge" is defined as $\theta_\text{open}$, while $\theta_\text{v}$ is the angle between the symmetry axis and the line of sight. In Fig. \ref{fig:geometry}, we provide a sketch illustrating the ejecta geometry.  
As the ejecta expands, a forward shock is driven into the ISM, and a reverse shock is driven into the ejecta.

\subsection{Coordinate systems}
The natural coordinate system for the ejecta initial conditions is a spherical coordinate system $(R,\phi,\theta)$, where $R$ is the distance from the origin, $\phi$ is the azimuthal angle, and $\theta$ is the polar angle, where the pole is aligned with the ejecta symmetry axis (see Fig. \ref{fig:geometry}). To calculate the synchrotron emission, it is convenient to define a coordinate system that is oriented with the line of sight. We define another spherical coordinate system, $(R,\varphi,\xi)$, where $\varphi$ is the azimuthal angle (chosen such that $\varphi=0$ corresponds to the equatorial plane in case of $\theta_\text{v}=\pi/2$), $\xi$ is the polar angle, and the pole is aligned with the line of sight. The coordinate transformation is 
\begin{equation}
\begin{aligned}
    \label{eq:coordinates}\cos\xi&=\cos\theta_\text{v}\cos\theta-\sin{\theta_\text{v}}\sin\theta\sin\phi,\\
        \sin\varphi&=\frac{\cos\theta-\cos\theta_\text{v}\cos\xi}{\sin\xi\sin\theta_\text{v}},
\end{aligned}
\end{equation}
similar to \cite{govreen-segal_analytic_2023}, although one difference is the choice of $\phi$ such that for $\phi=\pm\pi/2$ we have $\xi=\theta_\text{v}\pm\theta$.

A cylindrical coordinate system, $(r,\varphi,z)$, such that the $z$-axis is aligned with the line of sight, is useful for the semi-analytic calculation presented below. For such a coordinate system, the coordinate transformation is
\begin{equation}
\begin{aligned}
R&=\sqrt{r^2+z^2},\\
    \tan\theta &= \frac{\sqrt{(r\cos{\varphi})^2+(r\sin{\varphi}\cos{\theta_\text{v}}-z\sin{\theta_\text{v}})^2}}{r\sin{\varphi}\sin{\theta_\text{v}}+z\cos{\theta_\text{v}}}.
\end{aligned}
\end{equation}
\section{Semi-analytic calculation}
\label{sec:semi}
A semi-analytic approach for calculating the expected synchrotron emission from a toroidal ejecta with a velocity profile given by Eq. (\ref{eq:profile}) is presented below. This approach is valid as long as the reverse shock propagates through the ejecta steep, fast tail. After the reverse shock crosses the fast tail, a full 2D calculation is needed for a reliable calculation of the synchrotron emission.  
The semi-analytic analysis code offers a computation method for the approximated dynamics, anticipated flux, and the radio image on the sky.
\subsection{Dynamics}
\label{sec:dynamics}
The toroidal ejecta has mass, $M=M_{\text{iso}}\sin{\theta_\text{open}}$, where $M_{\text{iso}}$ is the isotropic equivalent mass
and $\theta_\text{open}$ is the ejecta opening angle.
The ejecta propagates through uniform ISM density, forming a forward-reverse shock structure, as shown in Fig.~\ref{fig:geometry}. 
As long as the reverse shock did not cross the steep fast tail, the total internal energy (and thus, the emissivity) of the system is dominated by the freshly shocked material within the initial opening angle because of the following argument: Given the steep ejecta profile with a power-law index of $s_\text{ft} \geq 5$, the energy and momentum injected into the shocked medium increase rapidly with decreasing velocity. This is because the mass of the ejecta, which scales as $M(>\gamma\beta) \propto(\gamma\beta)^{-s_\text{ft}}$, grows more quickly as velocity decreases. Consequently, the freshly shocked material, which is continuously energized by this injection, dominates the internal energy of the system. As the shock front propagates radially through the interstellar medium (ISM), the swept-up ISM mass is primarily composed of this freshly shocked material.
The unshocked ejecta, confined within an opening angle $\theta_\text{open}$, also moves radially, injecting additional energy and momentum into the shocked plasma. Due to conservation of momentum, the momentum of the freshly shocked material is dictated by this added momentum from the ejecta
at the position of the reverse shock.
We conclude the shocked plasma lateral expansion should have a limited effect on the observed emission before the reverse shock crosses the steep, fast tail. For larger values of $s_\text{ft}$, this effect becomes more pronounced, resulting in a less substantial lateral expansion.
Thus, we approximate the ejecta deceleration dynamics using the dynamics of stratified spherical ejecta with a corresponding mass of $M_{\text{iso}}=M/(\sin{\theta_\text{open}})$.

\subsection{Semi-analytic spherical dynamics}
\label{sec:semi_dynamics}
In \cite{sadeh_non-thermal_2023}, we developed a semi-analytical calculation method for modeling the dynamics of spherical mildly-relativistic ejecta with a velocity profile given by Eq. (\ref{eq:profile}), propagating into a cold and uniform ISM particularly applicable when the reverse shock propagates through the ejecta fast tail. Within the framework of this calculation, we approximated both shocked ISM and shocked ejecta as two uniform separate layers between the forward-reverse shock structure. 
This method is consistent with the full 1D relativistic hydrodynamics simulations \citep{sadeh_non-thermal_2023}.
The detailed derivation of this approach is presented in Appendix \ref{app:semi}.

\subsection{Intensity and flux}
\label{sec:semi_flux}
Using the approximated dynamics mentioned above, we obtain the flow profiles behind the shocks front: the internal energy density $e(R,t)$, the proper mass density $\rho'(R,t)$, and the Lorentz factor $\gamma(R,t)$, all as a function of the shock radius and time in the rest frame of the ISM. 
We approximate the shocked plasma as two separate uniform layers behind the shocks with different flow profiles for the two different layers. The thickness of these layers is determined by conservation of mass both for forward and reverse shocks. 
We employ a phenomenological approach to describe synchrotron emissivity: the fraction of energy carried by electrons is denoted by $\varepsilon_e$, and the fraction of energy carried by magnetic fields is denoted by $\varepsilon_B$. We assume a power-law electron distribution within the shocked layers, $dn_e/d\gamma_e\propto \gamma_e^{-p}$.
Following the assumption that the emission is dominated by shocked plasma within the initial opening angle, we restrict the emissivity to within $\theta_\text{open}$. Mathematically, this is expressed as:
\begin{equation}
j'_\nu=
\begin{cases}
j'_\nu \left(e,\rho,\gamma\right) & \text{for}\quad \frac{\pi}{2} -\theta_\text{open}<\theta<\frac{\pi}{2} +\theta_\text{open},\\
0 & \text{else},
\end{cases}
\end{equation}
where $j'_\nu \left(e,\rho,\gamma\right)$ is the emissivity as a function of the flow profiles, and the prime ($'$) notation indicates proper frame quantities. 
Subsequently, the emissivity and intensity are computed by the methodology outlined in \cite{sadeh_non-thermal_2024-1}: $j_\nu$ is defined by following \citet{rybicki_radiative_1979}, and $I_\nu$ is calculated by integrating over the $j_\nu$ while taking into account arrival time effects and relativistic effects. We only consider frequencies above the self-absorption frequency and below the cooling frequency, $\nu_a<\nu<\nu_c$, since the bulk of the radio to X-ray observations are expected to be in this part of the spectrum \citep{sadeh_non-thermal_2024-1}.
To calculate the flux, we consider the fact that the azimuthal observer symmetry breaks in the case of $\theta_\text{v}\neq 0$, so $I_\nu$ varies for different $\varphi$ values.
We integrate over the contribution from each azimuthal angle by
\begin{equation}
\begin{aligned}
    F_\nu&=\frac{1}{D^2}\int_0 ^R \int_0^{2\pi} r I_\nu drd\varphi,
    \end{aligned}
\end{equation}
where $R$ is the shock radius.
For the light curve temporal cutoff, we use the peak time estimation from \cite{sadeh_non-thermal_2023},
\begin{equation}
\label{eq:peak_t}
    t_\text{peak}= 550g(\beta_0)\left(\frac{M_{0,-4}}{n_{-2}}\right)^{\frac{1}{3}}\text{days},\quad
   g(\beta_0)=\frac{1.5-\sqrt{0.25+2\beta_0^2}}{\gamma_0^{\frac{1}{3}}\beta_{0}},
\end{equation}
where $M_0$ here is the isotropic equivalent mass of the fast tail ($M_{\text{iso}}=M/(\sin{\theta_\text{open}})$), $n=10^{-2}n_{-2}{\rm cm}^{-3}$ and $M_0=10^{-4}M_{0,-4}M_\odot$. For $0^\circ\leq\theta_\text{v}\lesssim20^\circ$ a delay is expected between the peak time in the isotropic equivalent case given by Eq. (\ref{eq:peak_t}) and the toroidal ejecta peak time due to lateral expansion of the shocked plasma after the reverse shock crosses the fast tail. This is due to relativistic beaming as $\sin^{-1}(1/\gamma_0)\sim70^\circ$ for typical values of $\gamma_0\approx1.05$. The peak flux in the isotropic equivalent case is proportional to $M_0$ \citep{sadeh_non-thermal_2023}. Since $\theta_\text{open}\sim15^\circ-30^\circ$ we expect the peak flux in the toroidal case to be lower by a factor of $\sin{15^\circ}-\sin{30^\circ}\sim2-3$ compared to the isotropic equivalent case. Thus, the peak flux is approximately \citep{sadeh_non-thermal_2023}
\begin{equation}
\begin{aligned}
    \label{eq:peak_f}
    F_{\nu,\rm peak}&\approx\sin\theta_\text{open}F^\text{iso}_{\nu,\rm peak},\\
    &\approx 4D_{26.5}^{-2}\varepsilon_{e,-1}^{p-1}f_{\rm ft}(p)
    \varepsilon_{B,-2}^{\frac{p+1}{4}}n_{-2}^{\frac{p+1}{4}}M_{0,-4}\nu_{9.5}^\frac{1-p}{2}g_q \mu\text{Jy},
    \end{aligned}
\end{equation}
with $D=10^{26.5}D_{26.5}$cm, $\nu=10^{9.5}\nu_{9.5}$Hz, $\varepsilon_{e}=10^{-1}\varepsilon_{e,-1}$ $\varepsilon_{B}=10^{-2}\varepsilon_{B,-2}$,
\begin{equation}
\begin{aligned}
\label{eq:gfactor}
g_q\approx1.2(\gamma_0\beta_0)^{2.2p-0.5},
\end{aligned}
\end{equation}
\begin{equation}
    f_{\rm ft}(p)=960\times8.8^{-p}(2.5-0.7p)(p-1)^{2-p}(l_p)^{p-1},
\end{equation}
\begin{equation}\label{eq:lp}
  l_p=\frac{p-2}{1-(\gamma_\text{max}/\gamma_\text{min})^{2-p}},\quad
  l_p\xrightarrow[p \to 2]{}\frac{1}{\ln(\gamma_\text{max}/\gamma_\text{min})},
\end{equation}
and $F^\text{iso}_{\nu,\rm peak}$ is the peak flux in the equivalent isotropic case. $f_{\rm ft}(p)$ deviates from unity by less than $30\%$ for $2<p<2.5$ \citep{sadeh_non-thermal_2023}.
\section{Numerical Calculation}
\label{sec:numerical}
\subsection{Numerical set-up}
\begin{figure}
    \centering
    \includegraphics[width=\linewidth]{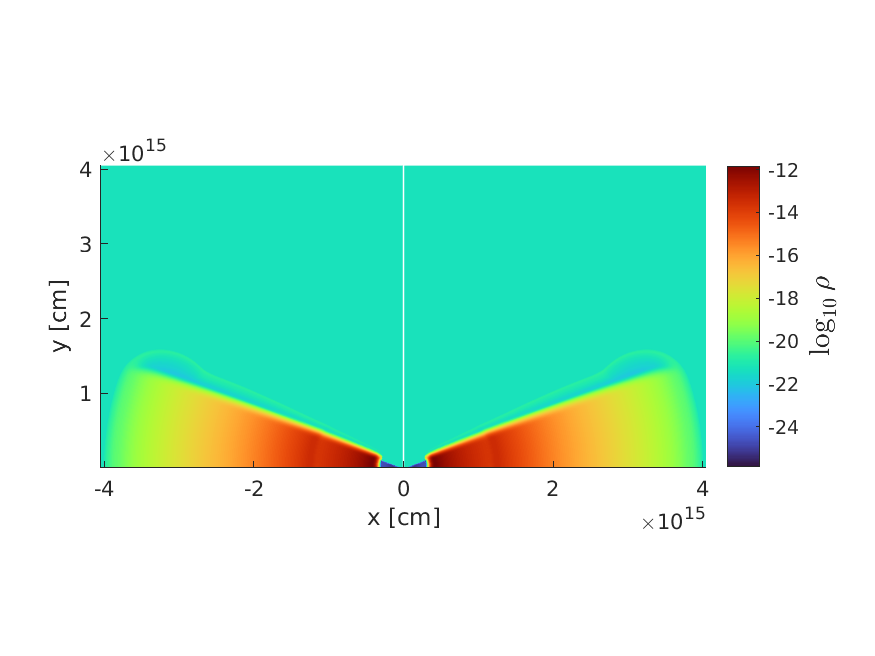}
    \caption{An example of a 2D density profile of a toroidal outflow propagating into a uniform interstellar medium calculated with RELDAFNA (the azimuthal symmetry axis is located at $x=0$). The initial time of the calculation is $10^5$s, and the snapshot is taken at $t=1.35\times10^5$s. The density scale is logarithmic. The initial conditions of the outflow are $\{s_\text{ft},s_\text{KN},\beta_0,n,M_R,\theta_\text{open}\}=\{6,1.5,0.3,3\times10^{2}\text{cm}^{-3},10^{-4}M_\odot,20^\circ\}$.}
\label{fig:snapshot}
\end{figure}

We employed numerical solutions of the 2D special relativistic hydrodynamics equations to validate our model through the RELDAFNA code \citep{klein_construction_2023}. RELDAFNA is an Eulerian code employing the Godunov scheme, incorporating adaptive mesh refinement (AMR) and second-order accuracy in time and space integration. Its efficient parallelization allows for high-resolution calculations, even for complex multiscale problems. RELDAFNA's accuracy was confirmed by comparing it with similar codes \citep{zhang_ram_2006, meliani_amrvac_2007} using standard test problems. Utilizing RELDAFNA, we solved the equations governing a relativistic outflow of an ideal fluid with a smoothly varying adiabatic index (between $5/3$ to $4/3$) depending on the internal energy density 
\begin{equation}
\hat{\gamma}=\frac{4+\left(1+1.3\frac{e}{\rho' c^2}\right)^{-1}}{3},
\end{equation}
which is in very good agreement with the EoS of mildly relativistic fluid provided by \cite{synge_relativistic_1957}. The $1.3$ factor is added for better agreement with the exact solution in the range of $0<e/\rho' c^2<2$ (corresponding to $1<\gamma<3$); see Appendix \ref{app:EOS}. Our simulations were performed in a 2D cylindrical coordinate system $(r,z)$ assuming axial symmetry, and we adopted a Courant–Friedrichs–Lewy (CFL) number of $0.5$, which provided efficient convergence (physical justified as long as it is $\leq1$).
The initial grid prioritizes resolving the ejecta by placing a higher concentration of cells within it, with a coarser spacing outside. Typically, the simulation begins with $500\times500$ cells, focusing most cells within the ejecta. AMR dynamically adjusts the resolution throughout the simulation, optimizing cell distribution along both the $r$ and $z$ axes to effectively capture regions of significant variation in pressure, density, or Lorentz factor. The initial pressure in all of the cells is set to $P=10^{-10}\rho_{\text{ISM}} c^2$ (where $\rho_{\text{ISM}}$ is the ISM mass density), and the computational domain boundaries are set at $r=10^{18}$cm and $z=10^{18}$cm.
We systematically refined the initial spatial grid and temporal steps to assess convergence.
We also iteratively increased the number of times the regridding scheme is allowed to multiply the number of cells within a time step. This process continues until no significant changes are observed in the hydrodynamics and emergent flux. Reflective boundary conditions were implemented at both the symmetry axis ($r=0$) and the symmetry plane ($z=0$) to account for symmetry considerations. 
Our numerical calculations adopt initial conditions comprising toroid ejecta, as illustrated in Fig. \ref{fig:geometry}. The ejecta is characterized by a flow profile described in Eq. (\ref{eq:profile}). 
The initial radius (at $t=t_0$) of a mass shell $M(>\gamma\beta)$ is determined by $r_0=\beta ct_0$. The expanding ejecta is initially embedded at $r>c t_0$ in a static, uniform, cold gas with negligible pressure. See Fig. \ref{fig:snapshot} for an example of a density map snapshot calculated with RELDAFNA.

\subsection{Numerical light curve}
From the 2D grid obtained via RELDAFNA, we construct a 3D grid by subdividing cells into segments that represent different azimuthal angles such that the number of segments is appropriate for achieving flux convergence. By post-processing the hydrodynamic profiles, we calculate specific emissivity, $j'_\nu(\vec{r},t)$ \citep[following][Appendix B]{rybicki_radiative_1979,sadeh_non-thermal_2024-1}, for frequency below the cooling frequency and above the synchrotron and self-absorption frequencies, $\nu_a,\nu_m<\nu<\nu_c$. The emissivity is calculated in the plasma proper frame for each cell within the 3D grid at a discrete set of lab times. Each fluid element is assumed to exist for a period of $\Delta t$ interval between time steps.
This calculation adopts a phenomenological approach for the magnetic field and electron energy distribution, assuming fractions $\varepsilon_B$ and $\varepsilon_e$ of internal energy density for magnetic fields and electrons, respectively. Furthermore, we assume a power-law energy distribution for electrons, $dn_e/d\gamma_e\propto \gamma_e^{-p}$.
The flux contribution from each cell for every observed time is computed as follows,
\begin{equation}
    \Delta F_\nu\left(t_\text{obs}=
    t-\frac{R\cos{\xi}}{c}\right)=\frac{j'_\nu(t)}{\gamma(t)^2(1-\beta(t)\cos{\xi_\beta})^2}\cdot\frac{\Delta V}{D^2},
\end{equation}
where $R$ is the distance from the origin, $\Delta V$ is the cell's volume, $D$ is the luminosity distance, $t$ is the time in the lab frame, $t_\text{obs}$ is the observer time, $\xi$ is the angle between the cell and the line of sight and $\xi_\beta$ is the angle between the cell velocity direction and the line of sight.
$\xi$ pose different values for different observation angles following Eq. (\ref{eq:coordinates}). Similarly, $\xi_\beta$ shares the same property by considering the following transformation, $\xi\rightarrow\xi_\beta,\theta\rightarrow\theta_\beta$ (where $\theta_\beta$ is the angle between the velocity direction and the symmetry axis), to Eq. (\ref{eq:coordinates}).
The total observed flux per unit frequency and observed time $t_\text{obs}$ is obtained by summing the different contributions $\Delta F_\nu$ from all of the fluid elements at all the discretized times.

\section{Validation of the semi-analytic method}
\label{sec:comparison}
In \S~\ref{sec:hydro}, we verify our estimation regarding the dominant contribution of the freshly shocked material within the initial opening angle of the ejecta to the internal energy. This verification holds true as long as the reverse shock hasn't yet crossed the ejecta fast tail. In \S~\ref{sec:light_curve}, we compare the light curves produced by the full 2D numerical calculation with our semi-analytic scheme to verify its validity. 
Finally, in \S~\ref{sec:map}, we compare the flux angular distribution obtained by our semi-analytic method to the full 1D numerical calculation of a spherical ejecta case to test its validity.
\subsection{Hydrodynamics}
\label{sec:hydro}
\begin{figure}
    \centering
\includegraphics[width=\columnwidth]{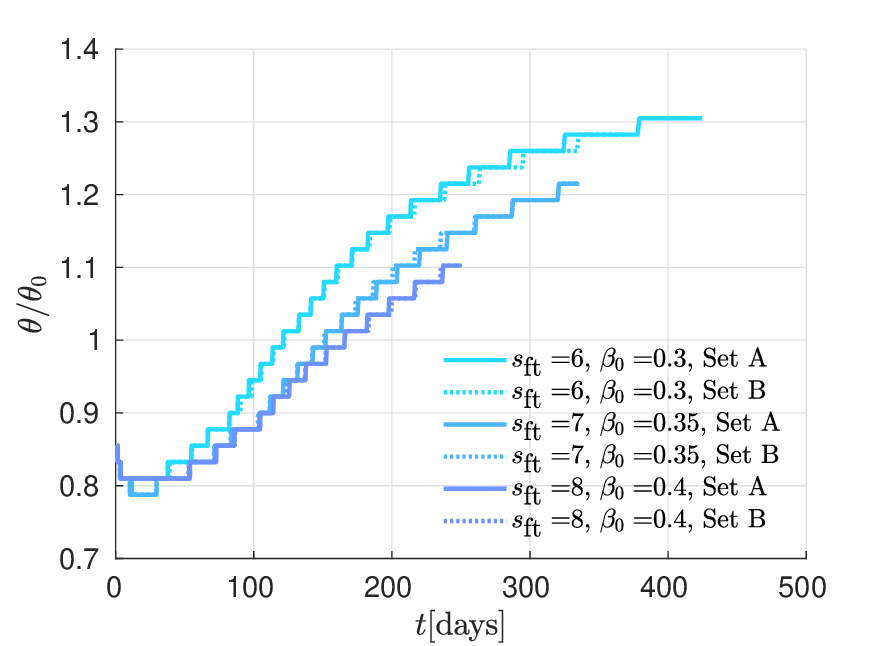}
    \caption{The opening angle (relative to the ejecta initial opening angle) within which $90\%$ of the internal energy, excluding rest mass energy, is included. 
    We consider few sets of toroidal ejecta parameters that affect the outflow dynamics: $\{s_\text{ft},\beta_0,\theta_\text{open}\}=\{6,0.3,20^\circ\}$ $/\{7,0.35,20^\circ\}$ $/\{8,0.4,20^\circ\}$, and $\{n,M_R\}=A: \{3\times10^{2}\text{cm}^{-3},10^{-4}M_\odot\}, B: \{3\text{cm}^{-3},10^{-6}M_\odot\}$. 
    The results are plotted for as long as the reverse shock propagates through the ejecta fast tail.}
    \label{fig:open_angle}
\end{figure}

 In our semi-analytic method, we assume that the internal energy is dominated by material within the initial opening angle of the ejecta. We ran numeric calculations with various initial conditions to test our assumption. The opening angle as a function of lab time is defined as the angle from the equatorial plane within which $90\%$ of the internal energy, excluding rest mass energy, is included. A few examples are given in Fig. \ref{fig:open_angle}. 
 We consider various parameter options since all of the following parameters $\{s_\text{ft},n,M_R,\}$ affect the ejecta dynamics, where $M_R\equiv M(\gamma\beta>1)= M_0(\gamma_0\beta_0)^{s_{\rm ft}}$,
is the "relativistic mass" with momentum $\gamma\beta>1$. We consider only typical values for the opening angle, $\theta_\text{open}=20^\circ$, and the velocity cutoff, $\beta_0=0.3-0.4$, that are consistent with the ejected material in merger simulations.
 We indeed find that the contribution to the internal energy is dominated by the material within the initial opening angle (up to $\sim 30\%$ corrections) as long as the reverse shock doesn't cross the ejecta fast tail. An evident trend in Fig. \ref{fig:open_angle} is that larger values of $s_\text{ft}$ lead to a less pronounced lateral expansion. This is because, at larger $s_\text{ft}$, the freshly shocked material makes up a larger fraction of the shocked mass, as explained in \S~\ref{sec:dynamics}.

\subsection{Light curves}
\label{sec:light_curve}
\begin{figure*}
    \centering
    \begin{subfigure}[b]{0.45\textwidth}
\includegraphics[width=\columnwidth]{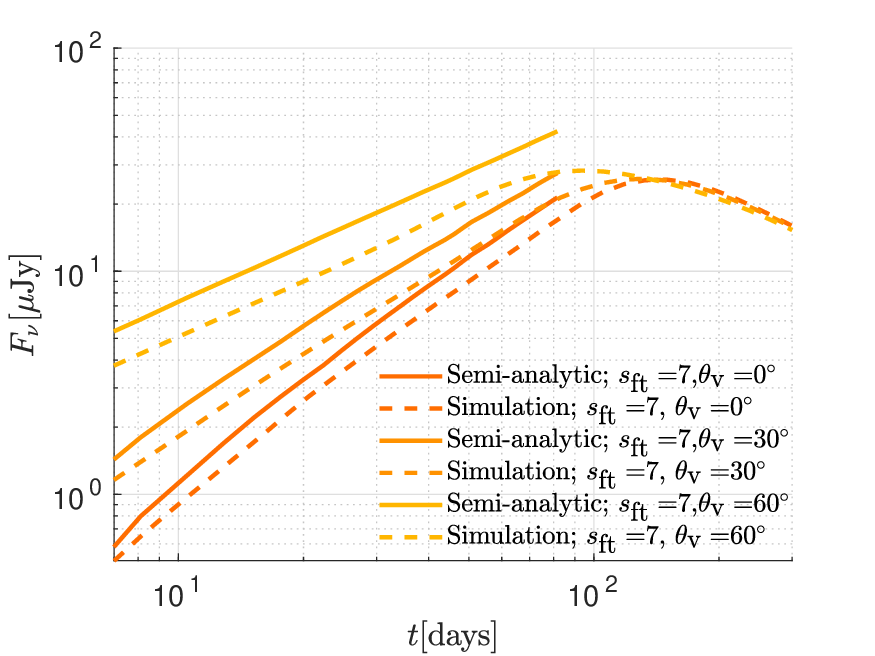}
    \end{subfigure}
    \hfill
\begin{subfigure}[b]{0.45\textwidth}
\includegraphics[width=\columnwidth]{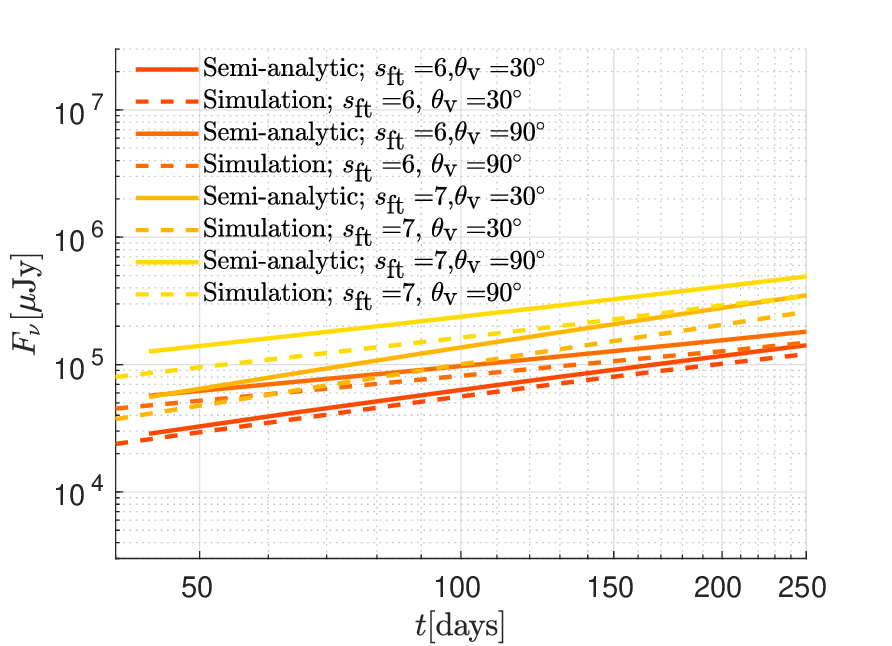}
\end{subfigure}
    \caption{A comparison between the flux calculated semi-analytically in (solid lines) and the flux calculated in the full numerical calculation (dashed lines), for various parameters. Left panel: $\{s_\text{ft},\beta_0,p,\theta_\text{v}\}=\{7,0.5,2.2,0^\circ/30^\circ/60^\circ\}$ and $\{n,M_R\}=\{3\text{cm}^{-3},10^{-6}M_\odot\}$. Right panel: $\{s_\text{ft},\beta_0,p,\theta_\text{v}\}=\{6,0.3,2.1,30^\circ/90^\circ\}$ $/\{7,0.35,2.2,30^\circ/90^\circ\}$ and $\{n,M_R\}=\{3\times10^{2}\text{cm}^{-3},10^{-4}M_\odot\}$. We also used $\{\varepsilon_e,\varepsilon_B,\theta_\text{open},s_\text{KN}\}=\{10^{-1},10^{-2},20^\circ,1.5\}$, a distance of $100$Mpc, and a frequency of $1$GHz.}
    \label{fig:flux}
\end{figure*}

We consider only frequencies that are between the self-absorption frequency/peak frequency to the cooling frequency, $\nu_a,\nu_m<\nu<\nu_c$, since the radio observations are expected to be in this part.
In Fig.~\ref{fig:flux}, we compare the light curve obtained semi-analytically (for as long as the reverse shock doesn't cross the fast tail, \S~\ref{sec:semi}) with the full 2D numerical calculation (\S~\ref{sec:numerical}).
The agreement is within 
$10$'s of percent for a range of relevant values of $\{n,M_R,\beta_0,s_{\rm ft},\theta_\text{v},\theta_\text{open},p\}$, along with $\varepsilon_e=10^{-1}$ and $\varepsilon_B=10^{-2}$, which modify the light curve based on a known analytical dependence, and $s_\text{KN}=1.5$, which alters the light curve after the peak. As expected, a $\sim20\%$ delay is observed for $\theta_\text{v}=0^\circ$.

\subsection{Sky image}
\label{sec:map}
\begin{figure}
    \centering
\includegraphics[width=\columnwidth]{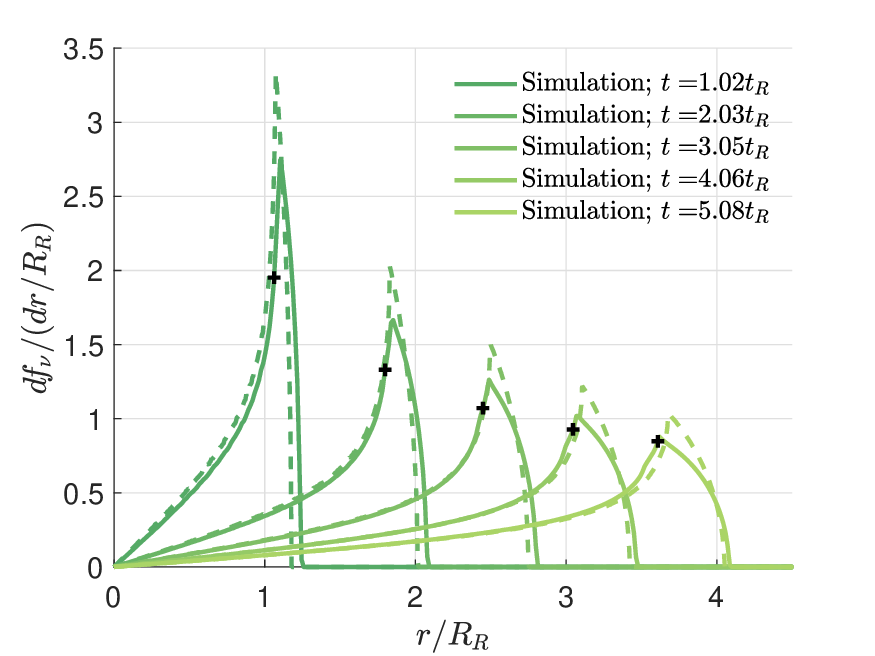}
    \caption{The fraction of flux that is emitted at different times by different annuli, $df_\nu/dr$, for $p=2.2,s_\text{ft}=7,\gamma_0=1.15$. 
    In solid lines: full numerical 1D calculation. In dashed lines: our semi-analytic scheme. $r$, the radius of the annulus (its transverse distance to the line of sight) is normalized to $R_R=ct_R$ (approximately the radius at which the post-shock plasma momentum drops to $\gamma\beta=1$), 
    where $t_R\equiv \left(\frac{M_R}{16 \pi nm_pc^3}\right)^\frac{1}{3}$. 
    The $+$'s are the analytic estimate of the image radius, from 
    \citet{sadeh_non-thermal_2024-1}.}
    \label{fig:ang}
\end{figure}

To further validate our semi-analytic calculation, in Fig.~\ref{fig:ang}, we show an example of the flux emitted from different annuli obtained numerically and semi-analytically for spherical ejecta (we considered $\theta_\text{open}=90^\circ$), along with the analytic estimate of the image radius from \citep{sadeh_non-thermal_2024-1}. Due to relativistic beaming and time arrival effects, the image is a relatively narrow ring. This ring radius increases with time since the blast wave expands and the relativistic beaming effect decreases. The emission from large viewing angles, which was initially negligible due to the relativistic beaming, becomes brighter as the velocity declines.

\section{Results}
\label{sec:results}
\begin{figure*}
    \centering
    \begin{subfigure}[b]{0.45\textwidth}
\includegraphics[width=\columnwidth]{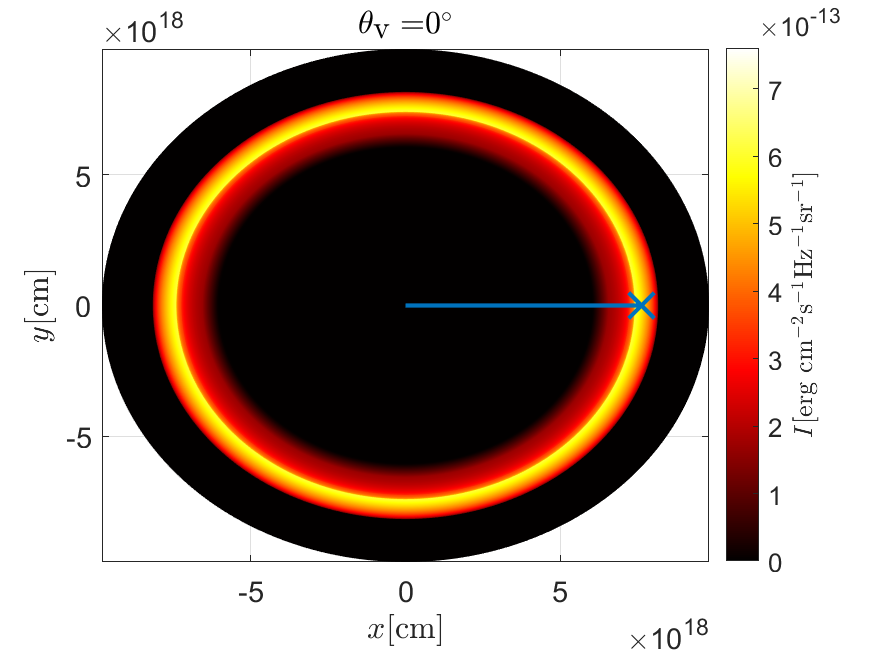}
    \end{subfigure}
    \hfill
\begin{subfigure}[b]{0.45\textwidth}
\includegraphics[width=\columnwidth]{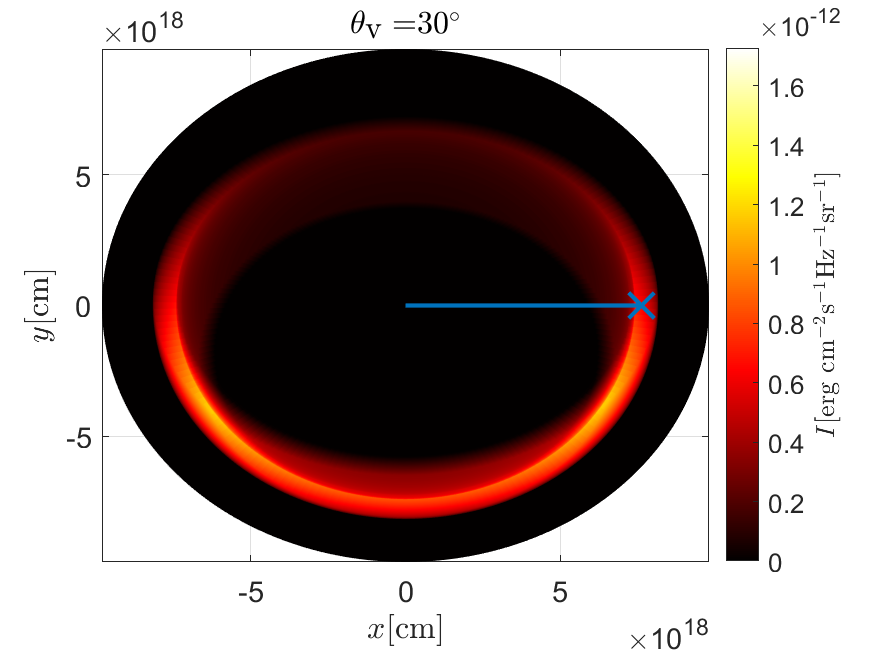}
\end{subfigure}
    \begin{subfigure}[b]{0.45\textwidth}
\includegraphics[width=\columnwidth]{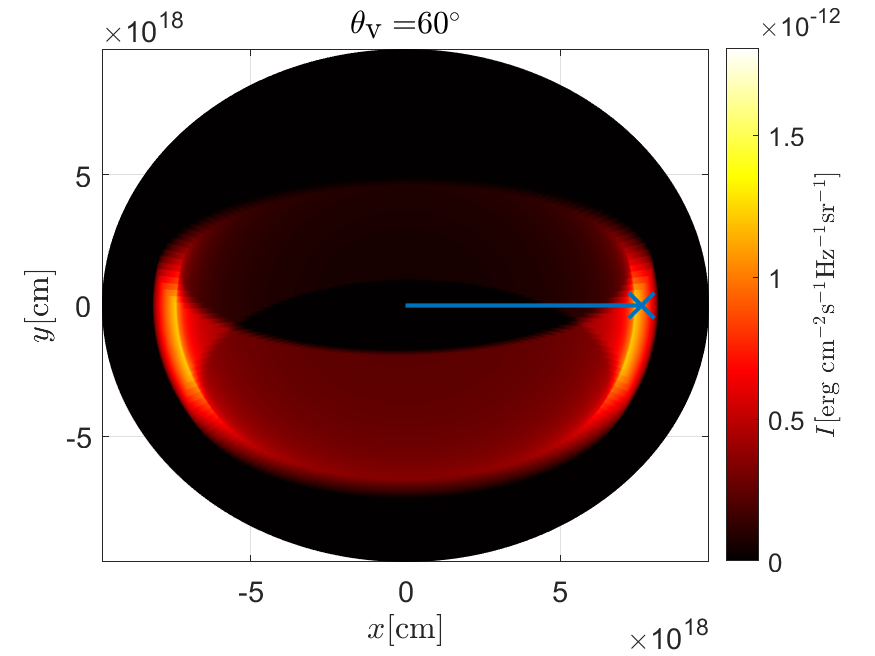}
    \end{subfigure}
    \hfill
\begin{subfigure}[b]{0.45\textwidth}
\includegraphics[width=\columnwidth]{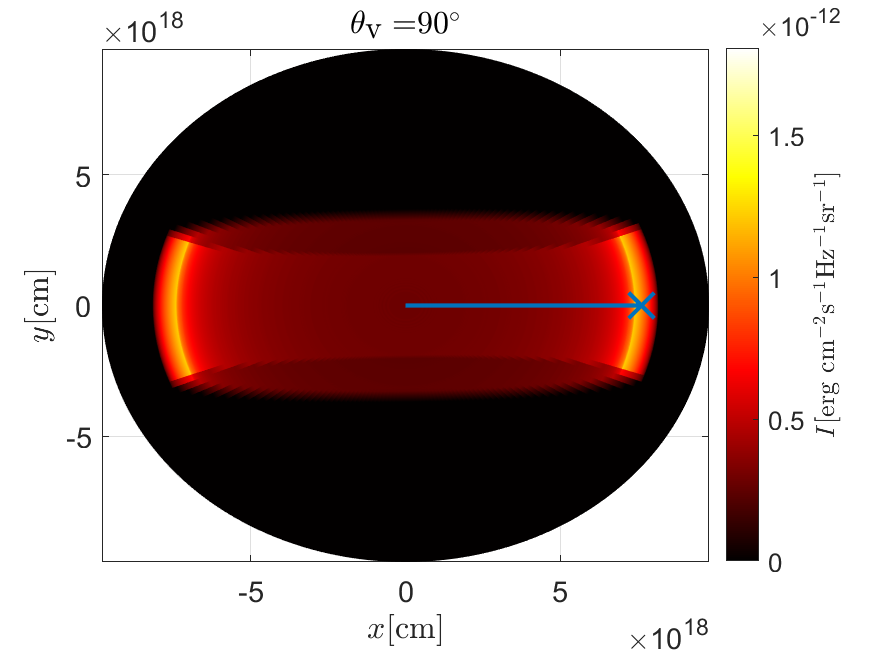}
\end{subfigure}
    \caption{The observed sky radio map for a distance of $100$Mpc, frequency of $1$GHz at $t=7110$days, for parameters expected for NSBH merger, $\{s_\text{ft},s_\text{KN},\beta_0,n,M_0,\varepsilon_e,\varepsilon_B,p,\theta_\text{open}\}=\{7,1.5,0.3,10^{-2}\text{cm}^{-3},8\times10^{-2}M_\odot,10^{-1},10^{-2},2.2,20^\circ\}$, and for different observation angles $\theta_\text{v}=0^\circ,30^\circ,60^\circ,90^\circ$, calculated by the semi-analytic method described in \S \ref{sec:semi}. At different times, the radius of the ring changes, but the physical picture remains.
    In blue: the analytic estimation from \citet{sadeh_non-thermal_2024-1} for observed image radius (the image appears as a narrow ring in the spherical case).}
    \label{fig:radio_maps}
\end{figure*}

\begin{figure}
\includegraphics[width=\columnwidth]{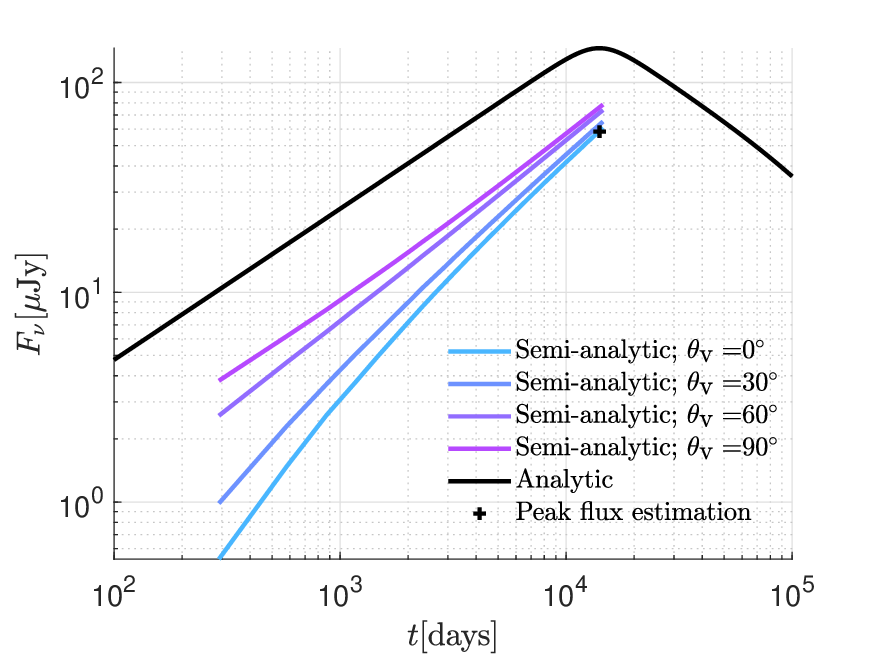}
    \caption{The observed light curves for a distance of $100$Mpc and a frequency of $1$GHz for parameters expected for NSBH merger, $\{s_\text{ft},s_\text{KN},\beta_0,n,M_0,\varepsilon_e,\varepsilon_B,p,\theta_\text{open}\}=\{7,1.5,0.3,10^{-2}\text{cm}^{-3},8\times10^{-2}M_\odot,10^{-1},10^{-2},2.2,20^\circ\}$, and for different observation angles $\theta_\text{v}=0^\circ,30^\circ,60^\circ,90^\circ$, calculated by the semi-analytic method described in \S \ref{sec:semi}. The black line represents the analytical model from \citet{sadeh_non-thermal_2023} for a spherical ejecta with the same isotropic equivalent mass. The black '+' is given by our estimations for the peak time from Eq. (\ref{eq:peak_t}) and peak flux from Eq. (\ref{eq:peak_f}).}
\label{fig:light_curves}
\end{figure}

The mass ejection in NSBH mergers simulations typically reaches $\sim0.1M_\odot$ in disruptive merger systems \citep{foucart_dynamical_2017,kyutoku_coalescence_2021,chen_black_2024}, implying an isotropic equivalent mass of $M\sim0.3M_\odot$ (for a typical opening angle of $\theta_\text{open}=20^\circ$). This is consistent with values of $M_R=10^{-4}M_\odot,s_\text{ft}=7$. 
In Figs.~\ref{fig:radio_maps}-\ref{fig:light_curves}, we provide several examples of the obtained radio maps and the corresponding light curves from the semi-analytic calculation for several values of $\theta_\text{v}$ and typical parameters expected for NSBH merger, $\{s_\text{ft},s_\text{KN},\beta_0,n,M_R,\varepsilon_e,\varepsilon_B,\theta_\text{open}\}=\{7,1.5,0.3,10^{-2}\text{cm}^{-3},10^{-4}M_\odot,10^{-1},10^{-2},20^\circ\}$. Notice that at the peak, the shocked plasma velocity is $\sim\beta_0=0.3$, and the radiation emitted from such plasma is beamed into an angle of $\sim\sin^{-1}(\frac{1}{\gamma_0})\approx 70^\circ$. This is the reason for the small difference at the peak flux between viewing angles of $0^\circ-30^\circ$, in which a larger fraction of the radiation is beamed away, to viewing angles of $60^\circ-90^\circ$. We add our analytic estimation for the light curve and image radius for a spherical ejecta case with the same equivalent isotropic mass for comparison.
We find that the toroidal ejecta case is less bright by a factor of $\sim2-3$ than the isotropic equivalent case (Fig. \ref{fig:light_curves}), and expected to peak on a time scale of ten years. Furthermore, the peak flux's dependence on the viewing angle is minimal for this toroidal ejecta. Oppositely, The radio map is significantly different between the different viewing angles (Fig. \ref{fig:radio_maps}), providing an opportunity to constrain the viewing angle in future radio map observations.
\section{Conclusions}
\label{sec:conclusions}
 In this work, we study the non-thermal emission of a toroidal (see Fig. \ref{fig:geometry}), mildly-relativistic ejecta with broken power-law mass distribution, given by Eq. (\ref{eq:profile}).
We provided a semi-analytic calculation method for the synchrotron emission, assuming emission only from the initial opening angle of the ejecta (\S \ref{sec:semi}) at times in which the reverse shock propagates through the steep, fast tail. Furthermore, we provided analytic estimations for the peak time and flux in Eqs. (\ref{eq:peak_t}) and Eq. (\ref{eq:peak_f}). Then, by calculating the full 2D hydrodynamics (\S \ref{sec:numerical}), we validated our semi-analytic approach in \S \ref{sec:comparison} for a wide range of ejecta parameter values characteristic of merger calculation results; an equatorial ejecta opening angle of $15^\circ<\theta<30^\circ$,
a steep ($s_{\rm ft}>5$, $\gamma_0\beta_0\geq0.3$) mass distribution for mildly relativistic velocities and a moderate mass distribution for the bulk of the ejecta ($1<s_{\rm KN}<3$, $0.1<\gamma_0\beta_0\leq0.3$). We also used various values of electrons power-law index, $2.1<p<2.4$, of the electron energy distribution.  
This is the first semi-analytic method supported by full 2D relativistic numeric calculations for the expected late non-thermal emission from mergers of compact objects with a substantial ($q>1.5$) mass ratio.
We found that the full 2D numerical calculation is consistent with our assumption (Fig. \ref{fig:open_angle}), i.e., that during reverse shock propagation through the fast tail, the freshly shocked material within the initial opening angle of the ejecta has a dominant contribution to the total internal energy.
Additionally, we found that, in agreement with Eq. (\ref{eq:peak_f}), the peak flux for toroidal ejecta is smaller by a factor of $\sim2-3$ (varies for different viewing angles) than spherical ejecta with the same isotropic equivalent mass, while the peak time estimation from (Eq. (\ref{eq:peak_t})) for spherical ejecta (with the same isotropic equivalent mass) is within $20\%$ deviation from the peak time results of the full numerical calculation for toroidal ejecta presented in this paper (Fig. \ref{fig:flux}). The radio map is significantly different between the toroidal and the spherical case in which the image is a symmetric ring (Fig. \ref{fig:radio_maps}). For $\theta_\text{v}\geq20^\circ$ the width of the observed image (for $\theta_\text{v}=90^\circ$ the width is parallel to the equatorial plane) is comparable with the ring diameter in the spherical case \citep{sadeh_non-thermal_2024-1}. Concurrently, the length (perpendicular to the width) is considerably smaller, providing a clear indication of equatorial ejecta. The different viewing angles affect the observed image orientation (see Fig. \ref{fig:radio_maps}), providing the possibility to constrain the viewing angle in future radio map observations.  
For $0^\circ\leq\theta_\text{v}\leq20^\circ$, the image is observed as a narrow ring (Fig. \ref{fig:radio_maps}), similar to the spherical case.
The semi-analytic scheme we presented in this work can be used to fit all of the model parameters $\{s_\text{ft},\gamma_0,n,M_R,\varepsilon_e,\varepsilon_B,p,\theta_\text{open},\theta_\text{v}\}$ to future observations of light curves and radio maps by an iterative process that scans the possible parameter space, e.g., Markov chain Monte Carlo. Degeneracies between the different parameters are expected to arise during the fitting process, for example, see Eq. (\ref{eq:peak_t})-(\ref{eq:peak_f}). There are two distinctions: the radio map shape is expected to provide stringent bounds over $\theta_\text{open}$ and $\theta_\text{v}$, while the synchrotron spectrum provides $p$.

Different mass ratios between the compact objects correspond to different ejecta geometries \citep{bernuzzi_accretion-induced_2020,bernuzzi_long-lived_2024}. Consequently, bounds over the ejecta geometry can constrain this mass ratio. For example, in GW170817, the estimation for the mass ratio varies between $1<q<2.6$ (for low spin assumption, it is $1<q<1.4$), while the estimation for the viewing angle from gravitational waves is $32^\circ\pm10^\circ$.
Observing the radio map of the late non-thermal emission, in this case, can potentially resolve the large uncertainty in the mass ratio.

The mass ejection in NSBH mergers simulations typically reaches $\sim0.1M_\odot$ in disruptive merger systems \citep{foucart_dynamical_2017,kyutoku_coalescence_2021,chen_black_2024}, implying an isotropic equivalent mass of $M\sim0.3M_\odot$ (for $\theta_\text{open}=20^\circ$). Following Eq. (\ref{eq:peak_f}), we expect these events to be brighter than the late non-thermal emission expected to follow BNS mergers, which typically have an isotropic equivalent mass of $M\sim 0.01M_\odot$ \citep{radice_binary_2018,nedora_dynamical_2021}, by a factor of $\sim5-10$.  
Thus, although the toroidal ejecta case is less bright by a factor of $\sim2-3$ than the isotropic equivalent case (see Fig. \ref{fig:light_curves}), the late non-thermal emission of NSBH merger is expected to be observed in radio at distances of $\lesssim200$Mpc by the Very Long Array (VLA) and expected to peak on a time scale of ten years for typical parameter values ($M_0=10^{-1}M_\odot$, $\beta_0=0.3$, $n=10^{-2}$cm$^{-3}$).
In conclusion, the peak flux's dependence on the viewing angle is minimal for this highly non-spherical geometry. Instead, among the ejecta parameters, the peak flux primarily relies on the total mass in the fast tail, as illustrated in Fig. \ref{fig:light_curves}. This suggests that our formula for the peak flux \citep{sadeh_non-thermal_2023} remains a valid estimate, within a factor of a $\sim$few, for the non-thermal peak flux from various 3D ejecta structures produced in full NR calculations of compact object mergers, provided that $M_0$ is replaced with the total mass in the fast tail.

\section*{Acknowledgements}
We thank Y. Klein and N.Wygoda for their authorization to use the RELDAFNA code and for their contribution. We also thank Noya Linder and Jonathan Morag for their helpful contribution. Finally, we thank Eli Waxman for helpful discussions and insightful comments.

\section*{Data Availability}
The data underlying this article will be shared following a reasonable request to the corresponding author.


\bibliographystyle{mnras}
\bibliography{references} 




\appendix
\section{Semi-analytic calculation description}
\label{app:semi}
\begin{figure}
\includegraphics[width=\columnwidth]{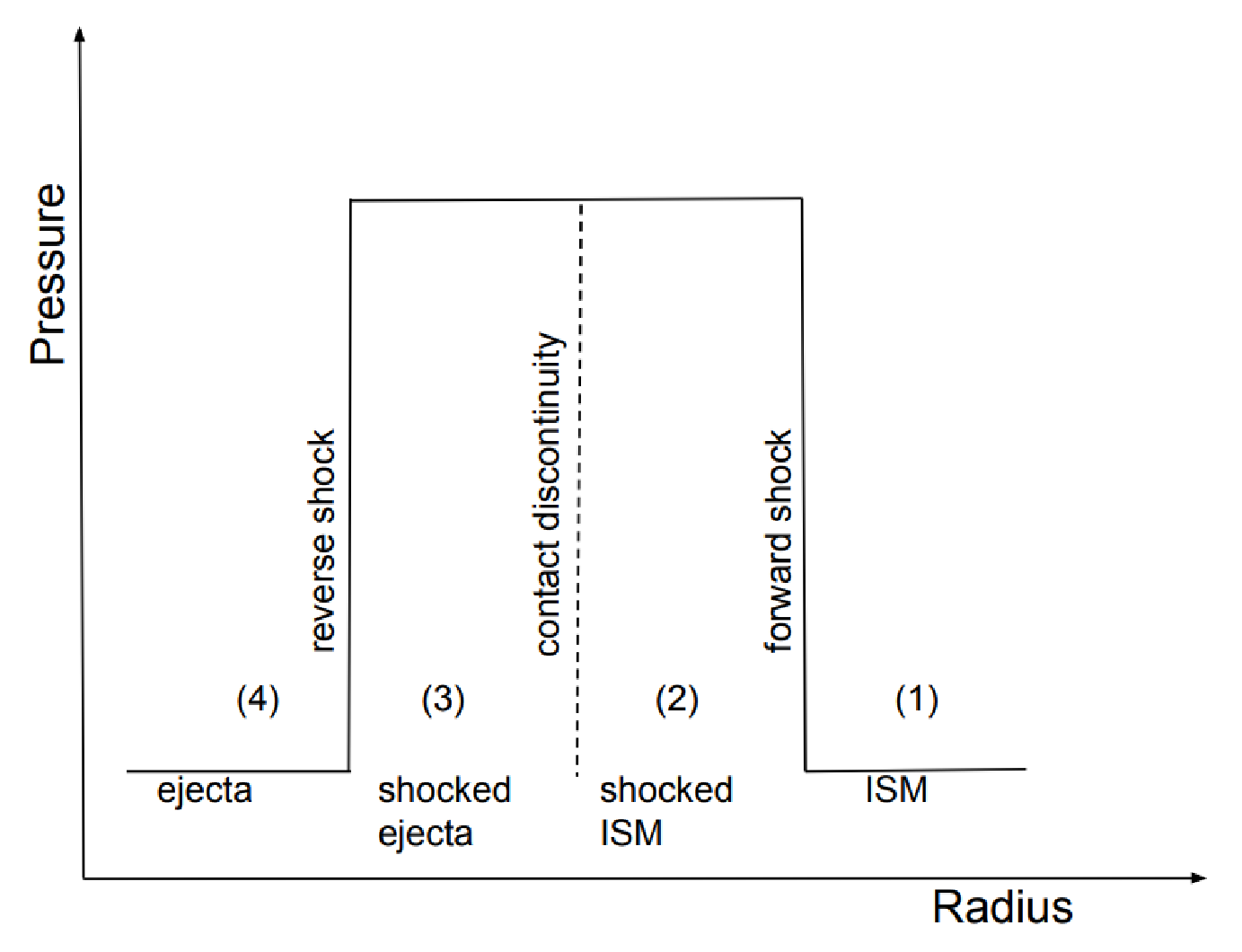}
    \caption{A schematic illustration showing the four regions of the forward-reverse shock structure described in Appendix} \ref{app:semi}: (1) un-shocked ISM, (2) shocked ISM, (3) shocked ejecta, (4) un-shocked ejecta. The pressure in the unshocked regions is negligible compared to that in the shocked regions, $p_1,p_4\ll p_2,p_3$. The pressure and velocity are approximated in our analytic calculations as uniform within the shocked region, and the density is approximated as uniform within the two regions separated by the contact discontinuity.
    \label{fig:FRshocks}
\end{figure}

As the ejected material expands, it initiates a forward shock that propagates into the ISM, while simultaneously, a reverse shock is generated within the ejected material itself (see Fig.~\ref{fig:FRshocks}). The deceleration imposed by the reverse shock reduces kinetic energy in the ejecta, which is subsequently transmitted to the heated ionized plasma within the shocked ISM.
We consider an ejecta with a mass profile given by $M(>\gamma_4\beta_4)=M_R(\gamma_4\beta_4)^{-s}$ (numeric subscripts relate to the region presented in Fig. \ref{fig:FRshocks}) and assume that the ISM density is uniform.
The initial energy of the ejecta (excluding the rest-mass energy) is given by
\begin{equation}
E(>\gamma_4\beta_4)=\int\frac{dM(>\gamma_4\beta_4)}{d\gamma_4}(\gamma_4-1)c^2d\gamma_4.
\end{equation}
As mentioned in $\S$ \ref{sec:semi} we approximate the shocked layers behind the shocks with uniform flow profiles ($e,\rho$ and $\gamma$), given by the shock jump conditions \citep{blandford_fluid_1976},
\begin{equation}
\label{eq:jump}
    \begin{aligned}
        e&=(\gamma-1)\rho'c^2,\\
        \rho'&=\frac{\hat{\gamma}\gamma+1}{\hat{\gamma}-1}\rho,
    \end{aligned}
\end{equation}
where $\rho'$ and $\rho$ are the mass density of the shocked material and the unshocked medium correspondingly, both in their own rest frame. $\hat{\gamma}$ is the adiabatic index relating the internal energy density, $e$, and the pressure, $p$ ($p=(\hat{\gamma}-1)e$). We use an adiabatic index $\hat{\gamma}(e/\rho')$ varying from $4/3$ to $5/3$ following the analysis of \cite{synge_relativistic_1957} for a plasma of protons and electrons. This is inaccurate for the shocked ejecta plasma, which is composed of a wide range of nuclei. However, the dependence of the results on the exact form of $\hat{\gamma}(e/\rho')$ for the shocked ejecta plasma is weak (as verified by the numeric calculations). 
We can relate the shocked ISM mass, $M_\text{ISM}$, and the shocked ejecta mass, $M(>\gamma_4\beta_4)$, to the mass densities in the rest frames by
\begin{equation}
\label{eq:dens}
    \begin{aligned}
        \rho_1&=\frac{3M_\text{ISM}}{4\pi R^3},\\
        \rho_4&=\frac{\gamma_4 s M(>\gamma_4\beta_4)}{4\pi (R_\text{cd}-\Delta_\text{ej})^3},
    \end{aligned}
\end{equation}
where $R$ is the forward shock radius, $R_\text{cd}$ is the contact discontinuity radius and $\Delta_\text{ej}$ the thickness of the shocked ejecta layer.
To find the shocked ISM thickness, $\Delta$, we consider mass conservation 
\begin{equation}
\begin{aligned}
    \frac{4\pi}{3}\rho_1R^3&=\frac{4\pi}{3}\rho_1\gamma_2\frac{\hat{\gamma}_2\gamma_2+1}{\hat{\gamma}_2-1}\left(R^3-(R-\Delta)^3\right),  \\
    \Delta&=R\left(1-\frac{\left(\gamma_2\frac{\hat{\gamma}_2\gamma_2+1}{\hat{\gamma}_2-1}-1\right)^\frac{1}{3}}{\left(\gamma_2\frac{\hat{\gamma}_2\gamma_2+1}{\hat{\gamma}_2-1}\right)^\frac{1}{3}}\right).
\end{aligned}
\end{equation}
The thickness of the shocked ejecta layer is given by
\begin{equation}
\begin{aligned}
    M(>\gamma_4\beta_4)&=\frac{4\pi}{3}\rho_4\gamma_2\frac{\hat{\gamma}_3\gamma_3^*+1}{\hat{\gamma}_3-1}\left(R_\text{cd}^3-(R_\text{cd}-\Delta_\text{ej})^3\right),  \\
\Delta_\text{ej}&=R_\text{cd}\left(1-\frac{\left(\frac{\gamma_2\gamma_4 s}{3}\cdot\frac{\hat{\gamma}_3\gamma_3^*+1}{\hat{\gamma}_3-1}\right)^\frac{1}{3}}{\left(1+\frac{\gamma_2\gamma_4 s}{3}\cdot\frac{\hat{\gamma}_3\gamma_3^*+1}{\hat{\gamma}_3-1}\right)^\frac{1}{3}}\right),
\end{aligned}
\end{equation}
where $\gamma_3^*=\gamma_2\gamma_4\left(1-\sqrt{1-\frac{1}{\gamma_2^2}-\frac{1}{\gamma_4^2}+\frac{1}{\gamma_2^2\gamma_4^2}}\right)$ is the Lorentz factor of region 3 in the frame of the unshocked ejecta.
Since $R-\Delta=R_\text{cd}$ we can write
\begin{equation}
\label{eq:r_del}
    (R_\text{cd}-\Delta_\text{ej})^3=R^3\left(1-\frac{\hat{\gamma}_2-1}{\gamma_2(\hat{\gamma}_2\gamma_2+1)}\right)\left(\frac{\left(\frac{\gamma_2\gamma_4 s}{3}\cdot\frac{\hat{\gamma}_3\gamma_3^*+1}{\hat{\gamma}_3-1}\right)^\frac{1}{3}}{\left(1+\frac{\gamma_2\gamma_4 s}{3}\cdot\frac{\hat{\gamma}_3\gamma_3^*+1}{\hat{\gamma}_3-1}\right)^\frac{1}{3}}\right).
\end{equation}
The total energy conservation equation, including pressure, is derived from the conservation of the energy-momentum tensor, $\partial_\mu T^{\mu0}=0$, such that the conserved quantity is \begin{equation}
    \int T^{00}dV=\int ((e+p)\gamma^2-p)dV.
\end{equation}
This can be expressed as
\begin{equation}
\label{eq:en_con}
    \begin{aligned}
        E(>\gamma_4\beta_4)&= E_{\text{k},2}+E_{\text{th},2}+E_{\text{k},3}+E_{\text{th},3},
    \end{aligned}
\end{equation}
where $E_{\text{k},i}/E_{\text{th},i}$ is the kinetic/thermal energy in region $i$. They are given by
\begin{equation}
    \begin{aligned}
        E_{\text{k},2}&=(\gamma_2-1)M_\text{ISM}c^2,\\
        E_{\text{th},2}&=(\gamma_2-1)\left(\gamma_2\hat{\gamma}_2-\frac{\hat{\gamma}_2-1}{\gamma_2}\right)M_\text{ISM}c^2,\\
        E_{\text{k},3}&=(\gamma_2-1)M(>\gamma_4\beta_4)c^2,\\
        E_{\text{th},3}&=(\gamma_3^*-1)\left(\gamma_2\hat{\gamma}_3-\frac{\hat{\gamma}_3-1}{\gamma_2}\right)M(>\gamma_4\beta_4)c^2.
    \end{aligned}
\end{equation}
Eq. (\ref{eq:en_con}) enables us to derive an expression for $M_\text{ISM}$ 
\begin{equation}
\label{eq:MISM}
    M_\text{ISM}=\frac{\frac{E(>\gamma_4\beta_4)}{c^2}-M(>\gamma_4\beta_4)\left(\gamma_2-1+(\gamma_3^*-1)\left(\gamma_2\hat{\gamma}_3+\frac{1-\hat{\gamma}_3}{\gamma_2}\right)\right)}{(\gamma_2-1)\left(1+\gamma_2\hat{\gamma}_2+\frac{1-\hat{\gamma}_2}{\gamma_2}\right)}.
\end{equation}
Finally, the equal pressure between regions 2 and 3, which are separated by the contact discontinuity, leads to
\begin{equation}
\label{eq:eq_p}
    (\hat{\gamma}_2\gamma_2+1)(\gamma_2-1)\rho_1c^2=(\hat{\gamma}_3\gamma_3^*+1)(\gamma_3^*-1)\rho_4c^2.
\end{equation}
We numerically solve Eq. (\ref{eq:eq_p}) by using the expressions in Eqs. (\ref{eq:dens}), (\ref{eq:r_del}) and (\ref{eq:MISM}) to find $\gamma_2(\gamma_4)$ for a given $\gamma_4$. Then we find the shock radius, $R(\gamma_4)$, and the flowfields, $\rho_3(\gamma_4),\rho_2(\gamma_4),e_3(\gamma_4),e_2(\gamma_4)$ by considering Eqs. (\ref{eq:jump}), (\ref{eq:dens}), and (\ref{eq:MISM}). 
The forward shock Lorentz factor, $\Gamma$ is given by \citep{blandford_fluid_1976}
\begin{equation}
\Gamma^2=\frac{(\gamma_2+1)(\hat{\gamma}_2(\gamma_2-1)+1)^2
}{\hat{\gamma}_2(2-\hat{\gamma}_2)(\gamma_2-1)+2}.
\end{equation}
Thus, the radius as a function of time, $R(t)$, can be found by numerically solving the following equation 
\begin{equation}
    \frac{dR}{dt}=\sqrt{1-\frac{1}{\Gamma^2}}.
\end{equation}
Using the above results, we finally define the shock radius, emitting layers thickness, and flow fields as a function of shock radius and/or lab time.
All calculations above are done in terms of $R_R,t_R$, and $M_R$ \citep{sadeh_non-thermal_2023} for better accuracy and efficiency. We multiply $R(t)$ by 0.95 and $\gamma_2\beta_2$ by 0.9 for better agreement with full numerical calculations.  

\section{Equation of State approximation}
\label{app:EOS}
\begin{figure}
    \centering
\includegraphics[width=\columnwidth]{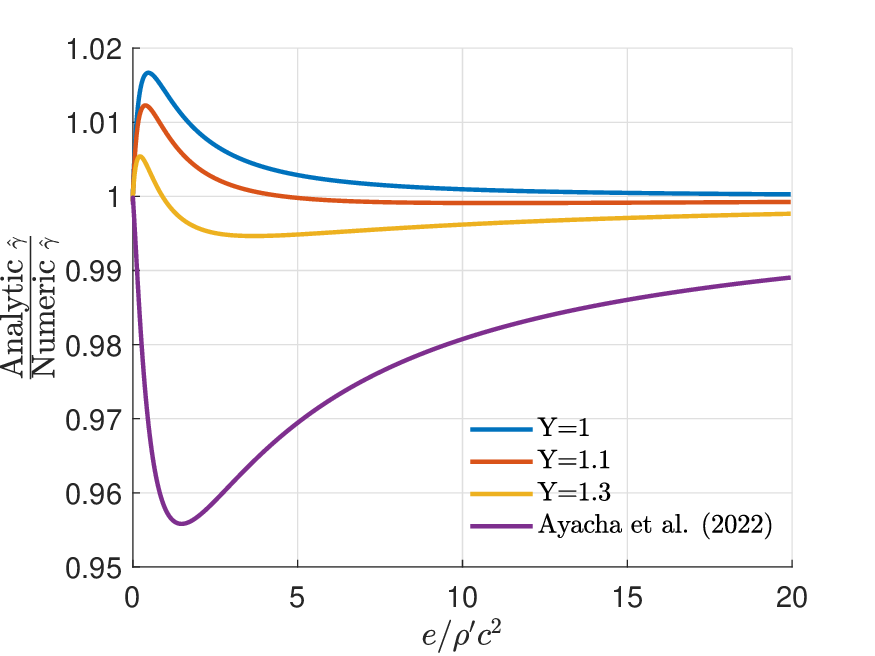}
 \caption{Comparison between the numerically calculated adiabatic index following \citet{synge_relativistic_1957} to our analytic approximation in Eq. (\ref{eq:EOS}), and to the analytic approximation suggested in \citet{ayache_gamma_2022}.}
    \label{fig:EOS}
\end{figure}

The theory of relativistic perfect gases gives us an expression for the specific enthalpy as a function of $\Theta\equiv \frac{p}{\rho' c^2}$,
that holds for a gas that is composed of the same particles and in the limit of a small free path when compared to the sound wavelength \citep{mignone_equation_2007}. It has the following form \citep{synge_relativistic_1957}
\begin{equation}
h=\frac{K_3(1/\Theta)}{K_2(1/\Theta)},
\end{equation} 
where $K_3$ and $K_2$ are, respectively, the modified Bessel functions of the second kind of order 3 and 2.
The specific enthalpy is defined by
\begin{equation}
    h=1+\frac{e}{\rho' c^2}+\frac{p}{\rho' c^2},
\end{equation}
plugging the relation $p=(\hat{\gamma}-1)e$ we obtain
\begin{equation}
\hat{\gamma}=\frac{h-1}{h-1-\Theta}=\frac{\frac{K_3(1/\Theta)}{K_2(1/\Theta)}-1}{\frac{K_3(1/\Theta)}{K_2(1/\Theta)}-1-\Theta},
\end{equation}
and 
\begin{equation}
\Theta=(\hat{\gamma}-1)\frac{e}{\rho' c^2}.
\end{equation}
These equations are solved numerically to find the adiabatic index of the fluid as a function of $\frac{e}{\rho' c^2}$.
In Fig. \ref{fig:EOS}, we compare this result with the following approximation 
\begin{equation}
\label{eq:EOS}
\hat{\gamma}=\frac{4+\left(1+Y\cdot\frac{e}{\rho' c^2}\right)^{-1}}{3},
\end{equation}
for different values of $Y=1/1.1/1.3$. We also compare the result to the approximation suggested in \cite{ayache_gamma_2022},
\begin{equation}
    \hat{\gamma}_\text{Ayacha}=\frac{4+\left(1+\frac{e}{\rho' c^2}\right)^{-2}}{3}.
\end{equation}
We find that our approximation fits well with the numerical results with the following $Y$ values,
\begin{equation}
    \begin{aligned}
    Y=\begin{cases}
            1,\quad \text{for } 9<\frac{e}{\rho'c^2},\\
            1.1,\quad \text{for } 2<\frac{e}{\rho'c^2}<9,\\
            1.3,\quad \text{for } 0<\frac{e}{\rho'c^2}<2.
        \end{cases}
    \end{aligned}
\end{equation}

\bsp	
\label{lastpage}
\end{document}